\documentclass[pra,amsmath,amssymb,twocolumn]{revtex4-1}

\usepackage{graphicx}
\usepackage[dvips]{epsfig}
\usepackage{bm}

\begin{document}

\title{Finite-range effects in dilute Fermi gases at unitarity} 

\author{Stefano Simonucci$^{*a,b}$, Giovanni Garberoglio$^{b}$, Simone
Taioli$^{b,c}$}

\affiliation{{\footnotesize $^{a}$ Department of Physics, University of Camerino,
via Madonna delle Carceri 9, 62032 Camerino, Italy - INFN Sezione Perugia}\\
{\footnotesize $^{b}$ Interdisciplinary Laboratory for Computational
Science (LISC), FBK-CMM and University of Trento, via Sommarive 18,
I-38123 Trento, Italy}\\
{\footnotesize $^{c}$ Department of Physics, University of Trento,
Via Sommarive 14, I-38123, Trento, Italy }}

\date{\today}
\begin{abstract}
We develop a theoretical method going beyond the contact-interaction
approximation frequently used in mean-field theories of many-fermion
systems, based on the low-energy $T$-matrix of the pair potential to
rigorously define the effective radius of the interaction. One of the
main consequences of our approach is the possibility to investigate
finite-density effects, which are outside the range of validity of
approximations based on $\delta$-like potentials. We apply our method
to the calculation of density dependent properties of an ultracold gas
of $^6$Li atoms at unitarity, whose two-body interaction potential is
calculated using {\em ab initio} quantum chemistry methods. We find
that density effects will be significant in ultracold gases with
densities one order of magnitude higher than those attained in current
experiments.
\end{abstract}

\maketitle

\section{Introduction}
\label{sec:intro}

In recent years, remarkable experimental advances in the field of trapped
alkali gases made it possible to tune the interaction of fermionic atoms and
study experimentally the transition between a BCS state of paired fermions to
a BEC state of diatomic molecules~\cite{Stringari}.

Several interesting phenomena occurring in these and other fermionic systems are usually analyzed
by using a mean-field treatment of a model short-range fermion-fermion
interaction, broadly assumed to be a contact potential
depending only on the s-wave scattering length $a$ of the form 
\begin{equation}\label{fci}
\hat{V}(\mathbf{r})= \frac{4\pi a}{m} \delta(\mathbf{r}).
\end{equation}  

This model potential, known as Fermi Contact Interaction (FCI), has
been used for calculating, for example, the weak interaction in
nucleon systems~\cite{Fermi} and for describing the Darwin correction
in the theory of hydrogen-like atoms~\cite{Heitler}.

FCI has been used successfully to describe Fermi systems (cold atomic
gases and nuclear matter in particular) in the mean-field
approximation, that is within the framework of the Bogoliubov--de
Gennes (BdG) equations
\cite{deGennes,Stringari,Gezerlis2011,Baksmaty,Bulgac2011}.  However,
the self-consistent solution of the BdG gap equation using FCI with an
external potential diverges, leading to an infinite
pairing~\cite{Randeria1}.  This divergence can be renormalized as
shown in Refs. \cite{Bulgac,Randeria1,Bruun}, even in the
experimentally and theoretically relevant unitary limit ($a\to
\infty$) \cite{EFT1,EFT2,cutoff}, where the behavior of a fermionic
system displays universal properties.  However, in this latter case,
the Hartree--Fock (HF) term appearing in the BdG equations diverges as
well, but this divergence is usually neglected
\cite{Stringari,Baksmaty}.

Recently, few research groups have investigated the effect of the
finite-range of the potential on the properties of dilute Fermi
systems, therefore going beyond the FCI description. In these
approaches the authors introduce an effective interatomic potential
characterized by two length scales which allow to fix independently
the scattering length and the potential range
\cite{Jensen04,DePalo04,Bruun05}.  Previous efforts to calculate the
energy spectrum of non ideal Fermi~\cite{Galitski} and Bose
~\cite{Beliaev} gases with short range potentials have been limited to
small densities by using perturbative expansion to the second order in
the parameter $k_Ff_0$, where $f_0$ is the scattering amplitude,
within Green's function formalism.  The importance of ultracold $^6$Li
gas in the unitary regime goes beyond the milestone experiment on
superfluid pairing by Ketterle~\cite{Ketterle} as it represents a
prototype of other strongly interacting fermions, such as neutron
matter~\cite{Gezerlis1} and dense quark matter~\cite{Alford}.

In this paper, we show how to include density effects in a mean-field
description of dilute Fermi gases. We achieve this goal by using the
on-shell $T$ matrix of the pair-potential as the appropriate
description of the interactions and of their effective range. We will
show that this approach removes the unphysical divergences obtained
with a $\delta$-like potential, and allows us to investigate -- in a
mean field approximation -- density effects in various observable
quantities characterizing dilute Fermi gases.  As a case-study, we
solve the BdG equations~\cite{deGennes,Pieri} for a homogeneous system
of $^6$Li atoms in the unitary regime, focusing on the definition of
the effective radius of the multichannel scattering potential,
obtained from {\em ab-initio} calculations.  At variance with
approaches based on the momentum expansion of the s-wave phase shift,
our theory predicts the same value of the potential range for both of
the Feshbach resonances present in this system.

This paper is organized as follows: in Sec.~\ref{sec:Tmatrix} we
derive the principal equations. We present our
approach where the contact potential is seen as a limit of separable
potentials, and the BdG equations are written in terms of the
$T$-matrix. This formalism is applied to ${}^6$Li in
Sec.~\ref{sec:abinit}, where we also describe how we obtained the
interatomic potential using {\em ab-initio} calculations. The results
regarding finite-range effects in this system are reported in
Sec.~\ref{sec:finite}.

\section{The $T$-matrix as the basic building block of BdG}
\label{sec:Tmatrix}

\subsection{The contact potential as a limit of separable potentials}

In general terms, one can write a two-body interaction potential as
\begin{eqnarray}
  \hat V &=& \int d\mathbf{r}_1 d\mathbf{r}_2 d\mathbf{r}'_1 d\mathbf{r}'_2 ~ 
\psi^\dagger(\mathbf{r}_1) \psi^\dagger(\mathbf{r}_2) \times \nonumber \\
& & V(\mathbf{r}_1, \mathbf{r}_2; \mathbf{r}'_1, \mathbf{r}'_2)
\psi(\mathbf{r}'_2) \psi(\mathbf{r}'_1)
\end{eqnarray}
where $\psi^\dagger(\mathbf{r})$ is the creation operator of a particle at
position $\mathbf{r}$.  If, the interaction is translationally invariant, then
the matrix elements in the previous equation factorize as:
\begin{eqnarray}
  V(\mathbf{r}_1,\mathbf{r}_2,\mathbf{r}'_1,\mathbf{r}'_2) &=& v(\mathbf{r}_1 - \mathbf{r}_2; 
  \mathbf{r}'_1 - \mathbf{r}'_2)  \times \nonumber \\
& & \delta \left(\frac{\mathbf{r}_1 + \mathbf{r}_2}{2} - \frac{\mathbf{r}'_1 + \mathbf{r}'_2}{2} \right).
\end{eqnarray}

Furthermore, a {\em local} potential is characterized by the condition
\begin{equation}
  v(\mathbf{r}; \mathbf{r'}) = \phi(\mathbf{r}) \delta(\mathbf{r-r'}),
\end{equation}
where $\phi(\mathbf{r})$ is the diagonal part of the potential and we have
defined $\mathbf{r} = \mathbf{r_1} - \mathbf{r_2}$ and $\mathbf{r'} = \mathbf{r'_1} - \mathbf{r'_2}$.  We call a potential
{\em separable} if there exists a function $f(\mathbf{r})$ so that
\begin{equation}
  v(\mathbf{r}; \mathbf{r'}) = f(\mathbf{r}) f(\mathbf{r'}).
\end{equation}

From the above definition sit is clear that the contact potential is
both local and separable. In the usual renormalization schemes, FCI is
obtained as the $n \to \infty$ limit of a a series of local potentials
$\phi_n(\mathbf{r})$.  In this paper we take the complementary route and define
the contact potential as a limit of separable potentials $f_n(\mathbf{r})$.  In
our approach the contact potential is seen as the limit of a series of
projectors onto a uni-dimensional manifold.
Therefore we write it as a limit of separable potentials with
progressively small radius:
\begin{equation}\label{vsep}
V(\mathbf{r},\mathbf{r}^{\prime})=\lim_{n\rightarrow\infty}f_{n}(\mathbf{r})V_{n}f_{n}^{*}(\mathbf{r}^{\prime})
\end{equation}
where $V_{n}$ is a term of this sequence (to be specified below) and
$f_{n}(\mathbf{r})$ is defined in term of a normalized arbitrary
function $f(\mathbf{r})$ such that
$f_{n}(\mathbf{r})=\sqrt{n^{3}}f(n\mathbf{r})$.  Our idea is to find
the sequence $V_n$ so that the $T$-matrix corresponding to the
potential in Eq.~(\ref{vsep}) describes the proper scattering length
$a$ of the system.  In doing so we will exploit the properties of the
on-shell $T$-matrix at zero energy. Its fundamental role in our
approach is further made clear by identifying the radius of the
effective range of the interactions using the expansion of the
on-shell $T$-matrix at zero momentum.  This definition of the
effective range differs from the usual one, which is based on the
value of the first order correction to the s-wave phase shift
\cite{Landau}
\begin{equation}\label{azero}
k~\cot\delta_0 = -\frac{1}{a}+\frac{1}{2} a_0 k^2 + \cdots
\end{equation}
In our approach we expand the on-shell $T$-matrix at zero energy, $T_{+}(0)$, around $\mathbf{k}=0$ 
obtaining:
\begin{equation}\label{expand}
T_{+}(0;\mathbf{k},\mathbf{k}^{\prime})\simeq\frac{4\pi a}{m}-c(k^{2}+k^{\prime2}).
\end{equation}

For $\mathbf{k},\mathbf{k}^{\prime}\approx 0$, $c$ is related to the range of the
on-shell $T_{+}(0)$ matrix in the $\mathbf{k}$-space or, by Fourier-transform, to the
inverse of the range of the $T_{+}(0)$ in the configuration space. Therefore,
we define the effective range of the potential as:
\begin{equation}\label{range}
r_{0}=\sqrt{\frac{cm}{\pi |a|}}.
\end{equation}

In order to specify the sequence $V_n$ appearing in Eq.~(\ref{vsep})
we remind that the scattering length is related to the on-shell
$T$-matrix via the equation:
\begin{equation}\label{adef}
\frac{4\pi a}{m}=\lim_{k\to 0} \langle k |T_+|k\rangle 
\end{equation}
and that the Lippmann--Schwinger (LS) equation $T = V+ V G_0(E) T$ can be
written as
\begin{equation}\label{tlip}
T(E) = \frac{1}{V^{-1} -G_0(E)},
\end{equation}
where $G_0(E)$ is the free-particle Green's function at energy $E$.
Expanding $G_0(E)$ in the momentum space one obtains:
\begin{equation}\label{eq1}
\langle f|\frac{1}{E-K+i\varepsilon}|f \rangle = 
Q_{0} + iQ_{1}k + \frac{1}{2}Q_{2}k^{2} + \cdots
\end{equation}
where $K$ is the kinetic energy, $E=\frac{k^{2}}{m}$, and it can
be shown that $Q_{1}=- m |\langle k=0|f \rangle|^{2} / (4 \pi)$.  
In terms of the elements of the sequence $f_{n}$ defined above, one has:
\begin{eqnarray}
\langle f_{n}|\frac{1}{E-K+i\varepsilon}|f_{n} \rangle &=& 
\frac{1}{n^{2}} \langle f|\frac{1}{\frac{E}{n^{2}}-K+i\varepsilon}|f \rangle \nonumber \\
&=&
\frac{Q_{0}}{n^{2}} + 
iQ_{1} \frac{k}{n^{3}}+ 
\frac{1}{2}Q_{2}\frac{k^{2}}{n^{4}} + \cdots
\end{eqnarray}
and therefore Eq.~(\ref{tlip}) becomes 
\begin{eqnarray}
\langle \mathbf{k}|T_{n+}|\mathbf{k}^{\prime} \rangle &=& \nonumber \\ 
&=& \frac{\langle {\mathbf{k}}/{n}|f \rangle 
          \langle f|{\mathbf{k}^{\prime}}/{n}\rangle / n^3}
{\left(
  \frac{1}{V_{n}} -
  \frac{Q_{0}}{n^{2}} - 
  iQ_{1}\frac{k}{n^{3}}-
  \frac{1}{2}Q_{2}\frac{k^{2}}{n^{4}}-\cdots
\right)}.
\end{eqnarray}

By choosing 
\begin{equation}\label{defpot}
V_{n}=\frac{n^{2}}{Q_{0}+\frac{Q_1}{an}}
\end{equation}
one obtains the correct scattering length in the limit $n\to\infty$ since, for $\mathbf{k},\mathbf{k}^{\prime}\approx 0$: 
\begin{eqnarray}\label{tuo}
\lim_{n\rightarrow\infty}
\langle \mathbf{k}|T_{n+}(0)|\mathbf{k}^{\prime} \rangle &=&
\frac{1}{n^{3}}
\frac{\langle {\mathbf{k}}/{n}|f\rangle \langle
  f|{\mathbf{k}^{\prime}}/{n}\rangle}
     {{1}/{V_{n}}-{Q_{0}}/{n^{2}}}
=\frac{4\pi a}{m}
\end{eqnarray}
With the same choice, the limiting value of the matrix elements of the potential are:
\begin{eqnarray}
\langle \mathbf{k}|V|\mathbf{k'}\rangle= \lim_{n\rightarrow\infty}
\langle \mathbf{k}|f_n\rangle V_{n}\langle f_n|\mathbf{k}^{\prime} \rangle &=& \nonumber \\
\lim_{n\rightarrow\infty}
\frac{\langle \mathbf{k}|f_{n}\rangle ~ n^{2} ~ \langle
  f_{n}|\mathbf{k}^{\prime}\rangle}
    {Q_{0}+\frac{Q_1}{an}}
 &=& \nonumber\\ 
\lim_{n\rightarrow\infty}\frac{1}{n^{3}}
\frac{\langle \mathbf{k}/n|f \rangle ~ n^{2} ~ 
      \langle f| \mathbf{k}^{\prime}/n \rangle}
     {Q_{0}+\frac{Q_1}{an}}
&=& 0.\label{mio}
\end{eqnarray}
Therefore, within our approach $T_{+}(0)$ is a $\delta$ function
for regular functions and has a different
behavior for not regular ones. 
At variance with Huang pseudo-potential, for any $g(\mathbf{r}) \sim 1/r$  at the
origin one obtains, by using the definition in Eq.~(\ref{defpot}):
\begin{eqnarray}
\lim_{n\rightarrow\infty}
\langle \mathbf{k}|V_{n}|g \rangle &=& \nonumber \\
\lim_{n\rightarrow\infty}
\frac{1}{n^{3}} 
\frac{\langle \mathbf{k}/n|f \rangle ~ n^{3} \langle f|g \rangle}
     {Q_{0}+\frac{Q_1}{an}}
 &=&\frac{\langle 0|f\rangle \langle f|g\rangle}{Q_{0}}
\end{eqnarray}
while, as by Eq.~(\ref{mio}), for regular functions the potential matrix
elements are zero.  

Notice that, from Eqs.~(\ref{vsep}), (\ref{adef}), and (\ref{tlip}) we
have, denoting by $| 0 \rangle$ the zero-momentum state and setting
$g_n = \langle f_n | G_0 | f_n \rangle$,
\begin{equation}
  \frac{4 \pi a}{m} = \lim_{n \to \infty} 
  \langle 0 | f_n \rangle 
  \frac{1}{n^3 \left( V^{-1}_n - g_n \right)}
  \langle f_n | 0 \rangle.
\end{equation}

In the $n \to \infty$ limit, the sequence of $\sqrt{n^3} \langle x | f_n \rangle$
is proportional to the $\delta$ function, and the matrix element $n^3 g_n$ tends to
infinity. As a consequence, one has
\begin{equation}
  \lim_{n \to \infty} \frac{V_n}{n^3} = 0.
\label{eq:limV}
\end{equation}

\subsection{The Bogoliubov--de~Gennes equations for a separable potential}

The Hamiltonian of a system of fermions in an external potential $\hat
U(x)$ and interacting with a two-body potential $\hat V$ is, in
the second quantized formalism,
\begin{equation} 
  H = \sum_{ij} \left( T_{ij} + U_{ij} \right) ~
               a^\dagger_i a_{j} + \frac{1}{2}
      {\sum_{il,jm}}  V_{il,jm} ~
      a^\dagger_{i} a^\dagger_{l} a_{m} a_{j},
\label{eq:H}
\end{equation}
where $a^\dagger_i$ and $a_i$ are the creation and annihilation
operators for a complete set of single particle states. The indices of these operators 
describe all the relevant quantum numbers.  

A mean-field solution of the ground state of the
Hamiltonian~(\ref{eq:H}) can be obtained by introducing an effective
Hamiltonian $H_\mathrm{eff}$, where the two-body potential $V$ in
Eq.~(\ref{eq:H}) is substituted by an effective one-body
potential
\begin{equation}
  V_\mathrm{eff} = \sum_{ij} \left(W_{ij} ~ a^\dagger_i a_j +
\frac{1}{2} \Delta_{ij}^* ~ a_i a_j + 
\frac{1}{2} \Delta_{ij} ~ a^\dagger_i a^\dagger_j \right),
\label{eq:Veff}
\end{equation}
where the as-yet unspecified matrices $W_{ij}$ and $\Delta_{ij}$ are
determined by requiring that the average values of $H$ and
$H_\mathrm{eff} - \mu \hat N$ are as close as possible to each other.
In the evaluation of the average value of $H$, one uses the
Hartree--Fock--Gorkov (HFG)
factorization of the two-body density matrix which, in the case of
fermions, is~\cite{Gorkov58}
\begin{equation}
  \langle a^\dagger_i a^\dagger_{l} a_{m} a_{j} \rangle = 
  \langle a^\dagger_i a_j \rangle \langle a^\dagger_l a_m \rangle -
  \langle a^\dagger_i a_m \rangle \langle a^\dagger_l a_j \rangle +
  \langle a^\dagger_i a^\dagger_l \rangle \langle a_j a_m \rangle.
\label{eq:Gorkov}
\end{equation}

The self-consistent equations for $W_{ij}$ and $\Delta_{ij}$ turn out
to be:
\begin{eqnarray}
  W_{ij} &=& \sum_{l,m} 
  \left( V_{il,jm} - V_{il,mj} \right) 
  ~ \langle a^\dagger_l a_m \rangle \label{eq:W} \\
  \Delta_{ij} &=& -\sum_{l,m} V_{ij,lm} ~ \langle a_l a_m \rangle,
  \label{eq:Delta}
\end{eqnarray}
and one is left with a two-body effective Hamiltonian, which can be put in the
form
\begin{equation}
H_\mathrm{eff} = \sum_k \epsilon_k ~ b^\dagger_k b_k + E_0,
\label{eq:Heff}
\end{equation}
where $E_0$ is the ground state energy. The sum over $k$ is
restricted to those states where $\epsilon_k \geq 0$. The new set of
fermionic operators $b_k$ and $b^\dagger_k$ are given by the
Bogoliubov transform
\begin{equation}
  \begin{array}{rcl}
  a_i &=& \displaystyle\sum_j \left( u_{ij} b_j + v_{ij}^* b^\dagger_j \right) \\
  a^\dagger_i &=& \displaystyle\sum_j \left( u^*_{ij} b^\dagger_j + v_{ij} b_j
  \right),
  \end{array}
\label{eq:bogoliubov}
\end{equation}
and satisfy the relation
\begin{equation}
  \langle b^\dagger_i b_j \rangle =
  \frac{\delta_{ij}}{\exp\left(\displaystyle\frac{\epsilon_i}{k_B
      T}\right) + 1},
\label{eq:bexpval}
\end{equation}
although in the following we will be concerned, for the sake of
conciseness, with the $T \to 0$ limit.

The equations determining the coefficients $u_{ij}$ and $v_{ij}$ appearing in
equation~(\ref{eq:bogoliubov}) are the well known Bogoliubov--de~Gennes
equations~\cite{deGennes}:
\begin{equation}
  \begin{array}{rcl}
  \displaystyle\sum_k \left[ \left( T_{ik} + U_{ik} + W_{ik} - \mu \delta_{ik}
    \right) u_{kj} + \Delta_{ik} v_{kj} \right] &=& \epsilon_j u_{ij}, \\
  \displaystyle\sum_k \left[ \left( T_{ik} + U_{ik} + W_{ik} + \mu \delta_{ik}
    \right)^* v_{kj} + \Delta^*_{ik} u_{kj} \right] &=& - \epsilon_j
  v_{ij},    
  \end{array}
\label{eq:bdg}
\end{equation}
which have to be solved self-consistently with the definitions of
$W_{ij}$ and $\Delta_{ij}$ given in equations~(\ref{eq:W}) and
(\ref{eq:Delta}), respectively.

In our approach, where the FCI is seen as a limit of separable
potentials, the HF term of Eq.~(\ref{eq:W}) turns out to be zero. In
order to see that, we remind that the Latin indices in the previous
equations were a short-hand notation to indicate both spin and spatial
degrees of freedom. For example, one has
\begin{equation}
  \langle a^\dagger_l a_m \rangle = \varrho_{\alpha \beta}(\mathbf{r}, \mathbf{r'})
  = \sum_{\eta} \int d\mathbf{x} ~ v_{\alpha \eta}(\mathbf{r},\mathbf{x}) v^*_{\beta \eta}(\mathbf{r'},\mathbf{x})
\end{equation}  
where $\varrho_{\alpha \beta}(\mathbf{r}, \mathbf{r'})$ is the one-body density matrix,
with the Greek letters indicating spin degrees of freedom.
If we further assume that the pair potential is given by
\begin{equation}
  V_{ij,lm} = \delta_{\alpha \gamma} \delta_{\beta \delta} 
  f_n(\mathbf{R}-\mathbf{R'}) V_n f_n(\mathbf{r}-\mathbf{r'}) 
  \delta\left(\frac{\mathbf{r}+\mathbf{r'}}{2} - \frac{\mathbf{R} + \mathbf{R'}}{2} \right),
\end{equation}
then the HF term is given by:
\begin{eqnarray}
  W_{\alpha \beta}(\mathbf{r}, \mathbf{r'}) &=& \lim_{n \to \infty}
  \frac{V_n}{n^3} \delta(\mathbf{r}-\mathbf{r'}) \sum_\eta \int d\mathbf{x} \times \nonumber \\
    & & \left[ 
      \delta_{\alpha \beta} \sum_\gamma 
      v_{\gamma \eta}(\mathbf{r},\mathbf{x}) v^*_{\gamma \eta}(\mathbf{r'},\mathbf{x})  - \right. \nonumber \\
    & & \left . v_{\alpha \eta}(\mathbf{r},\mathbf{x}) v^*_{\beta \eta}(\mathbf{r'},\mathbf{x}) \right],
\end{eqnarray}
from which, using Eq.~(\ref{eq:limV}), we obtain that the HF term is
identically zero. Notice that the result of this derivation does not
depend on whether the Fermi gas is free or confined.
This result also shows that neglecting the HF term in the BdG equations,
as usually done in the theoretical mean-field treatment of Fermi gases, is
indeed a consistent choice.

Moreover, these results allow one to cancel out divergences in the
self-consistent equations, e.g. in the pairing function, without ad-hoc
renormalization procedures, and clarify the short-range nature of the
potential in the sense of the limit to the $\delta$ function, and extend the
calculations to finite densities.

To test the ability of this finite-radius approach to solve the divergence
problem, we apply our scheme to the solution of the BdG equations for an
homogeneous system of strongly interacting fermions at unitarity.  The
solution for ultracold atoms at unitarity using FCI requires a nontrivial
renormalization procedure~\cite{Randeria1}. 

\subsection{Scattering view of the Bogoliubov--de~Gennes equations}

In our approach, we rewrite the BdG equations at zero temperature in
terms of the on-shell $T$-matrix and then perform our limit procedure,
retaining Eq.~(\ref{range}) as a definition of the effective range of
the interaction potential.

The equation for the pairing function $\Delta$ defined in
Eq.~(\ref{eq:Delta}) is readily seen to be
\begin{eqnarray}
  \Delta_{\alpha \beta}(\mathbf{r}, \mathbf{r'}) &=&
  - \int d\mathbf{R} d\mathbf{R'} ~ \sum_{\alpha' \beta'}
  V_{\alpha\beta, \alpha' \beta'}(\mathbf{r},\mathbf{r'};\mathbf{R},\mathbf{R'})
    \times \nonumber \\
    & & \sum_{\gamma} \int d\mathbf{x} ~ u_{\alpha' \gamma}(\mathbf{R},\mathbf{x}) v^*_{\beta' \gamma}(\mathbf{R'},\mathbf{x})
  \nonumber \\
  &\equiv&  -V {\cal Q} \label{eq:D=VQ}
\end{eqnarray}
where in the last equality we have formally written the double
integral as a ``matrix product''.  In the case of the FCI potential,
the equation as it stands plagued by a ultraviolet divergence.

Equation~(\ref{eq:D=VQ}) can be rewritten using the
Lippmann--Schwinger equation~(\ref{tlip}), so that the pairing
function is determined by the normalized interaction (embodied in the
$T$ matrix) instead of the ``bare'' interaction described by the
potential $V$\cite{Randeria1}.

This is achieved by rewriting the LS equation as $V = (1 - V G_0(E))
T$ and considering the quantity $\Delta - V G_0(E) \Delta$. Using
Eq.~(\ref{eq:D=VQ}), we have the equalities
\begin{eqnarray}
  \Delta - V G_0 \Delta &=& -V {\cal Q} - V G_0 \Delta \\
                        &=& -V ({\cal Q} + G_0 \Delta) \\
                        &=& -(1 - V G_0) T ({\cal Q} + G_0 \Delta)
\end{eqnarray}
from which, assuming that $(1 - V G_0(E))$ is invertible, we get
\begin{equation}
  \Delta = -T ({\cal Q} + G_0(E) \Delta)
\label{eq:Delta_dressed}
\end{equation}
which is an equation for the pairing function involving the $T$ matrix
instead of the ``bare'' potential $V$. The solution of
equation~(\ref{eq:Delta_dressed}) has of course to be determined self
consistently with the solution of the BdG equations~(\ref{eq:bdg}).
Introducing the indices $\alpha$ and $\beta$, denoting internal
degrees of freedom of the atoms, Eq.~(\ref{eq:Delta_dressed}) reads:
\begin{eqnarray}\label{self}
\Delta_{\alpha\beta,\mathbf{p}} =
-\int\frac{d\mathbf{p}^{\prime}}{(2\pi)^{3}} \sum_{\alpha' \beta'}
T_{+;\alpha\beta,\alpha^{\prime}\beta^{\prime}}(E;\mathbf{p},\mathbf{p}^{\prime})
\times \nonumber
\\ \left(\sum_{q}u_{\alpha^{\prime}q,\mathbf{p}^{\prime}}v_{\beta^{\prime}q,\mathbf{p}^{\prime}}^{*}+
\frac{1}{E-E_{\alpha^{\prime}}-E_{\beta'}-\frac{p^{\prime2}}{m}}
\Delta_{\alpha^{\prime}\beta^{\prime}, \mathbf{p}^{\prime}}\right).
\end{eqnarray}

In the previous expression the value of the parameter $E$ can be
chosen arbitrarily and we have used this freedom to have a better
convergence of the self-consistent solution.

\section{Ab-initio calculation of the multichannel $T$-matrix}
\label{sec:abinit}

We use Eq. (\ref{self}) to study an ultracold gas of $^6$Li atoms at
unitarity~\cite{Simonucci}, where the scattering length is infinite,
and the observables, notably the pairing gap and the chemical
potential, are proportional to the Fermi momentum of the free
gas~\cite{Stringari}. Such a regime can be obtained by exploiting the
Feshbach resonance of a $^6$Li spin-mixture.  Experimental
measurements~\cite{Jochim,Bartenstein2005} show two Feshbach
resonances for a magnetic strength of $543.28$~G (narrow resonance)
and $822 \cdots 834$~G (broad resonance), respectively.

The atom-atom collisions are described by an Hamiltonian of the form
\begin{equation}
  H = T + H_\mathrm{el} + V_\mathrm{hf} + V_B,
\end{equation}
where $T$ is the kinetic energy of the nuclei, $H_\mathrm{el}$ is the
electron Hamiltonian (including electronic kinetic energy,
Coulomb potential and spin-orbit coupling), and $V_\mathrm{hf}$ is the
hyperfine interaction. We have also included a term $V_B$ describing
the interaction of the atoms with an external magnetic field.

In the usual Born--Oppenheimer approximation, the wave equation for
the relative motion of a pair of ultracold atoms at zero angular
momentum can be written as
\begin{equation}
  - \frac{\hbar^2}{2 m} \nabla^2 \phi_\alpha(r) + 
  V_\alpha(r) \phi_\alpha(r) +
  \sum_\beta V_{I,\alpha \beta}(r) \phi_\beta(r)
 = E \phi_\alpha(r)
\label{eq:nuclei}
\end{equation}
where the label $\alpha$ denotes the total electron and nuclear spin
states of the colliding atoms. $V_\alpha(r)$ is the internuclear
potential obtained using the ground-state electronic wavefunction for
the given value of $\alpha$, and $V_{I,\alpha \beta}(r)$ describes the
coupling induced by the hyperfine interaction term, $V_\mathrm{hf}$,
and the interaction with the magnetic field, $V_B$.

From the solution of Eq.~(\ref{eq:nuclei}) in the $E \to 0$ limit, one
can obtain the s-wave scattering length $a$ of the system from the
asymptotic values of the wavefunction, that is
\begin{equation}
  \phi_\alpha(r) = A_0 \left(1 - \frac{a}{r} \right).
\end{equation}

\begin{figure}
\includegraphics[width=0.80\linewidth]{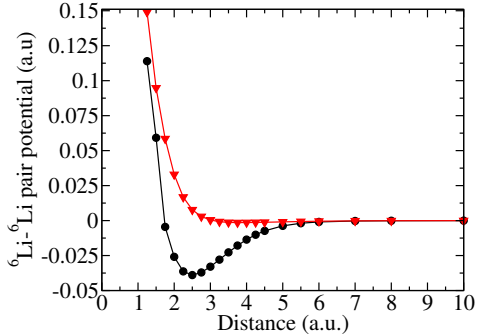}
\caption{\label{PES} PES for the singlet (dots) and
triplet (triangles) states in $^6$Li-$^6$Li scattering.}
\end{figure}

In our theoretical analysis, the $^6$Li-$^6$Li pair potentials
appearing in Eq.~(\ref{eq:nuclei}) as a function of the external
magnetic field have been calculated by using Configuration Interaction
with single and double excitations from the reference Hartree--Fock
ground state, taking into account fine and hyperfine structure
terms~\cite{PhyRep}.  Slater determinants have been built by expanding
the molecular orbitals in a cc-pVQZ basis set of atomic-centered
Gaussians.  The electronic singlet and triplet potential energy
surfaces are reported in Fig.~\ref{PES} using dots and triangles,
respectively.

\begin{figure}
\includegraphics[width=0.50\columnwidth,angle=-90]{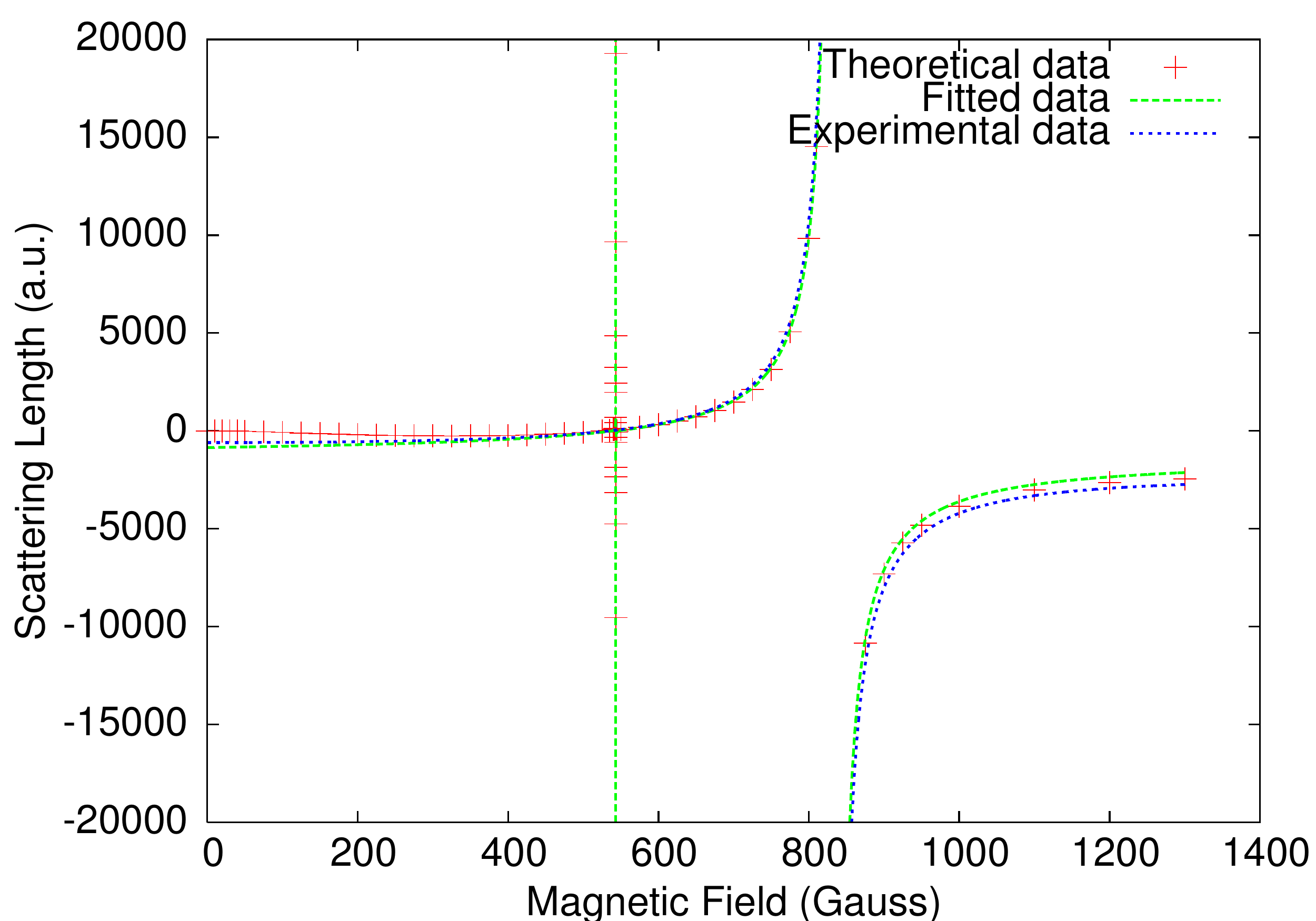}
\caption{\label{scatlength} S-wave scattering length vs applied magnetic
  field for the hyperspin states ($f,m_f$ ) = (1/2,-1/2) and 
(1/2, 1/2) in $^6$Li-$^6$Li scattering.}
\end{figure}

In Fig.~\ref{scatlength} we plot the calculated s-wave scattering
length vs magnetic field for the lowest energy hyperspin doublet open
channel.  The positions of the narrow ($543.25$~G) and broad ($834$~G)
resonances are in good agreement with the experimental
data~\cite{Bartenstein2005}.

The $T$-matrix elements appearing in Eq.~(\ref{self}) have been
computed by means of multichannel scattering theory~\cite{PhyRep}.  To
perform this calculation we have included relativistic terms (of the
order of $2 \times 10^{-6}$~a.u.), which mix the singlet and triplet
states, and we have multiplied each of the {\it ab-initio} curves of the
potential by a single parameter fixed to reproduce the values of the
Feshbach resonances.


Since the open and closed channels are coupled via the hyperfine
interaction, the problem is very complex if one takes into account the
full on-shell multichannel $T$-matrix in the gap
equation~(\ref{self}).  Therefore, while the calculation of $T_{+}(0)$
has been performed in the multichannel space, we projected the
$T$-matrix on the lowest hyperspin-doublet open channel to solve the
BdG gap equation. This assumption is less drastic than using the
interaction potential projected onto the open channel. In fact, the
projected multichannel $T_{+}(0)$ matrix retains short range
interactions with the closed channels and this is the reason why we
prefer to use $T_{+}(0)$ rather than $V$ in the self-consistent
equation (\ref{self}).

The on-shell $T_{+}(0)$ is represented over $10^5$ equally spaced grid
points, while a grid of $200\times200$ $k$-points according to
a Gauss--Chebyshev quadrature over the interval [0,1] has been used in
the momentum space, to obtain a convergence below $10^{-5}$ a.u.  At
unitarity, we find that the values of the effective range $a_0$, which
is usually defined in term of the s-wave phase shift (see
Eq.~(\ref{azero})) are $-600$~a.u. and $80$~a.u for the narrow and
broad resonances, respectively.  The dependence of $a_0$ on the resonance
position and its large negative value in the narrow resonance rule out
$a_{0}$ as a measure of the effective range of the screened potential
at unitarity.  

On the other hand, by using Eq.~(\ref{range}) as a definition of the
effective radius, we obtain for $r_0$ values of 26.5806~a.u. and
26.5756~a.u. for the narrow and broad resonance, respectively, with a
difference of less than 0.02~$\%$.
We consider this result as a further indication that the definition of
the effective range based on the small-momentum expansion of the
$T$-matrix is reasonable and gives consistent results.

Our methodology provides directly the values of the relevant
quantities characterizing an ultra-cold Fermi gas -- such as the gap,
the chemical potential or the transition temperature -- as a function
of the effective range $r_0$. Their behavior is discussed in the
following section.

\section{Finite-range effects in the unitary limit for $^6$Li}
\label{sec:finite}

\begin{figure}
\includegraphics[width=0.80\linewidth]{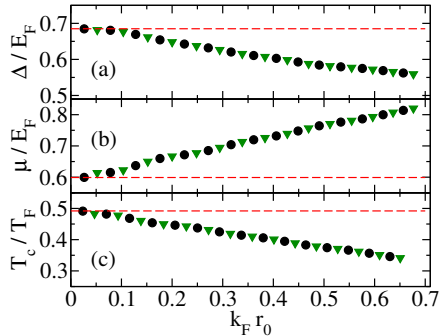}
\caption{\label{gapmu}
Dimensionless gap function (a), chemical potential
(b), and critical temperature (c) at unitarity for the narrow
(dots), broad (triangles) resonances and for the $\delta$-like interaction
(dashed lines).}
\end{figure}

The ratios $\mu/E_{F}$ and $\Delta/{E_{F}}$ at unitarity as a function
of $k_{F}r_{0}$ are plotted in the boxes (a) and (b) of
Fig.~\ref{gapmu}, respectively. We observe that the behavior is
exactly the same for the two resonances, as expected from the fact
that we obtain the same value of the effective range for both the
narrow and wide resonances.

These quantities are increasingly deviating from the curve of the
$\delta$-like interaction (dashed horizontal line) as the density is
increased.  

Using Quantum Monte Carlo (QMC) calculations, Gezerlis {\em et
  al.}~\cite{Gezerlis1,Gezerlis2,Schirotzek,Carlson} obtained
universal values at unitarity of $\Delta/E_f \approx 0.475 \pm 0.075$,
$\mu/E_f \approx0.4 \pm 0.075$, while Chang {\em et al.}~\cite{Chang}
and Bulgac {\em et al.}~\cite{Bulgac2} obtained $\Delta/E_f \approx
0.6 \pm 0.02$ and $ \Delta/E_f \approx 0.58 \pm 0.02$, $\mu/E_f
\approx 0.44 \pm 0.01$ and $\mu/E_f \approx 0.38 \pm 0.06$
respectively.  It is not surprising that our calculations yield
results that are consistently higher than those provided by QMC, since
the BdG equations are based on a mean-field approximation of the
fermion-fermion interaction. The BdG results are therefore very close
to the BCS limit of $\Delta/E_f \approx 0.68$ and $\mu/E_f \approx
0.6$.

Although the derivation of the main equations of our approach has been
carried out in the $T \to 0$ limit for the sake of conciseness, our
theory can be straightforwardly extended to finite temperatures using
Eq.~(\ref{eq:bexpval}). As a consequence, we can calculate the
transition temperature $T_c$ -- defined as $\Delta(T_c) = 0$ -- as a
function of the parameter $k r_0$.  This quantity, which is a very
important parameter for any superfluid system, is reported in box (c)
of Fig.~\ref{gapmu}. The ratio $T_c/T_f$ has been previously estimated
0.494 at unitarity for a homogeneous
system~\cite{Holland,Haussman}. Note that our theory predicts the same
transition temperature at unitarity for both resonances, and it
deviates up to 30~\% of the corresponding FCI value by increasing the
density of the system.

Our results show a significant density dependence for $\Delta$, $\mu$
and $T_c$, in contrast with the predictions obtained using FCI for
which a constant value is obtained for all these quantities.  We
estimate that the largest density that we have investigated
(corresponding to $k_F r_0=0.6$) is about one order of magnitude
larger than that achievable in modern harmonic
traps~\cite{Ketterle}. Nevertheless, this value is likely reachable by
experimental apparatus in a nearest future.  At this high density, one
might expect that three-body recombination could lead significant atom
losses.  However, we estimate that at equilibrium and for near
threshold conditions 3-body recombination will play a little role even
at such increased density regime, since the kinetic energy gain in the
recombination process ($10^{-17}$ a.u.)  is very small compared to the
Fermi energy ($8\times10^{-8}$ a.u.)  \cite{Suno}.

\section{Conclusions}
\label{sec:conclusions}

To summarize, we have proposed a new definition of contact potential based on
the the radius of the on-shell $T$-matrix at zero energy.  This analysis
underpins a short range, rather than a $\delta$-like, model of the screened
fermion-fermion interaction and rules out the scattering length and the usual
effective radius as relevant parameters to describe dilute quantum gases at
unitarity.  Furthermore, we have shown the ability of our approach to cancel
out naturally the divergences arising with the use of the contact potential.
Finally, the application of this new theoretical approach to the
self-consistent solution of the BdG equations for ultracold $^6$Li
homogeneous gas in the unitary regime has been discussed.

\acknowledgments

We thank Prof. Francesco Pederiva for useful discussions. S.T.
acknowledges economical support from Provincia Autonoma di Trento,
under the Marie-Curie Action ``People Programme'' of the 7th FP ``The
Trentino programme of research, training and mobility of post-doctoral
researchers -- Outgoing Researcher''.

\bibliographystyle{apsrev4-1}
\bibliography{pra}

\end{document}